\documentclass[prd,superscriptaddress,amsfonts,amssymb,amsmath,onecolumn]{revtex4}
\usepackage{bm}
\usepackage{amsfonts}
\usepackage{latexsym}
\usepackage{graphicx}
\usepackage{amsmath}
\usepackage{palatino}
\usepackage{mathpazo}
\usepackage{textcomp}
\usepackage{float}
\usepackage{booktabs}
\usepackage{dcolumn}
\usepackage{booktabs}
\usepackage{multirow}
\usepackage{hyperref}
\hypersetup{colorlinks,citecolor=blue}
\usepackage{amsmath}
\usepackage{xcolor}
\usepackage{orcidlink}
\usepackage[caption=false]{subfig}
\usepackage{commath}
\def\jnl@style{\it}
\def\aaref@jnl#1{{\jnl@style#1}}

\def\aaref@jnl#1{{\jnl@style#1}}

\def\aj{\aaref@jnl{AJ}}                   % Astronomical Journal
\def\apj{\aaref@jnl{ApJ}}                 % Astrophysical Journal
\def\apjl{\aaref@jnl{ApJ}}                % Astrophysical Journal, Letters
\def\apjs{\aaref@jnl{ApJS}}               % Astrophysical Journal, Supplement
\def\apss{\aaref@jnl{Ap\&SS}}             % Astrophysics and Space Science
\def\aap{\aaref@jnl{A\&A}}                % Astronomy and Astrophysics
\def\aapr{\aaref@jnl{A\&A~Rev.}}          % Astronomy and Astrophysics Reviews
\def\aaps{\aaref@jnl{A\&AS}}              % Astronomy and Astrophysics, Supplement
\def\mnras{\aaref@jnl{Mon.~Not.~Roy.~Astron.~Soc.}}             % Monthly Notices of the RAS
\def\prd{\aaref@jnl{Phys.~Rev.~D}}        % Physical Review D
\def\prc{\aaref@jnl{Phys.~Rev.~C}}  % Physical Review C
\def\prl{\aaref@jnl{Phys.~Rev.~Lett.}}    % Physical Review Letters
\def\qjras{\aaref@jnl{QJRAS}}             % Quarterly Journal of the RAS
\def\skytel{\aaref@jnl{S\&T}}             % Sky and Telescope
\def\ssr{\aaref@jnl{Space~Sci.~Rev.}}     % Space Science Reviews
\def\zap{\aaref@jnl{ZAp}}                 % Zeitschrift fuer Astrophysik
\def\nat{\aaref@jnl{Nature}}              % Nature
\def\aplett{\aaref@jnl{Astrophys.~Lett.}} % Astrophysics Letters
\def\apspr{\aaref@jnl{Astrophys.~Space~Phys.~Res.}} % Astrophysics Space Physics Research
\def\physrep{\aaref@jnl{Phys.~Rep.}}      % Physics Reports
\def\physscr{\aaref@jnl{Phys.~Scr}}       % Physica Scripta
\def\commat{\aaref@jnl{Comm.~Math.~Phys.}}              % Communications in Mathematical Physics
\def\science{\aaref@jnl{Science}}               % Science
\def\cqg{\aaref@jnl{Classical Quant.~Grav.}}            % Classical and Quantum Gravity
\def\jpcs{\aaref@jnl{JPCS}}                                     % Journal of Physics Conference Series
\def\ijmpd{\aaref@jnl{Int.~J.~Mod.~Phys.~D}}                    % International Journal of Modern Physics D
\def\grg{\aaref@jnl{Gen.~Relat.~Gravit.}}               % General Relativity and Gravitation
\def\rpp{\aaref@jnl{Rep.~Prog.~Phys.}}          % Reports on Progress in Physics
\def\npa{\aaref@jnl{Nucl.~Phys.~A}}        % Nuclear Physics A
\def\lrr{\aaref@jnl{Living Rev.~Rel.}}                   % Living reviews in relativity
\def\jcap{\aaref@jnl{J.~Cosmology Astropart.~Phys.}}    % Journal of cosmology and astroparticle physics
\def\rmp{\aaref@jnl{Rev.~Mod.~Phys.}}   %Reviews of modern physics
\def\epjc{\aaref@jnl{Eur.~Phys.~J.~C}}

\allowdisplaybreaks[1]
\begin{document}

\color{black}       %% For one column

\title{Thermodynamics of the Van der Waals black hole within nonextensive Kaniadiakis entropy}
%\title{Nonextensive Kaniadakis thermodynamics of the Van der Waals black hole}
%with a Chaplygin-like equation of state}

\author{Adam Z. Kaczmarek}
\email[Email: ]{adamzenonkaczmarek@gmail.com }
\affiliation{Institute of Physics, Faculty of Science and Technology, Jan D{\l}ugosz University in Cz{\c{e}}stochowa, 13/15 Armii Krajowej Ave., 42200 Cz{\c{e}}stochowa, Poland}
\author{Yassine Sekhmani\orcidlink{0000-0001-7448-4579}}
\email[Email: ]{sekhmaniyassine@gmail.com}
\affiliation{Center for Theoretical Physics, Khazar University, 41 Mehseti Street, Baku, AZ1096, Azerbaijan.}
\affiliation{D\'epartement de Physique, Equipe des Sciences de la mati\`ere
et du rayonnement, ESMaR, \\
\small Facult\'e des Sciences, Universit\'e Mohammed V de Rabat, Rabat,
Morocco}

\author{Dominik Szcz{\c{e}}{\'s}niak\orcidlink{0000-0003-1880-1255}}
\email[Email: ]{d.szczesniak@ujd.edu.pl}
\affiliation{Institute of Physics, Faculty of Science and Technology, Jan D{\l}ugosz University in Cz{\c{e}}stochowa, 13/15 Armii Krajowej Ave., 42200 Cz{\c{e}}stochowa, Poland}

\author{Javlon Rayimbaev\orcidlink{0000-0001-9293-1838}}
\email[Email: ]{javlon@astrin.uz}
\affiliation{Institute of Fundamental and Applied Research, National Research University TIIAME, Kori Niyoziy 39, Tashkent 100000, Uzbekistan}
\affiliation{University of Tashkent for Applied Sciences, Str. Gavhar 1, Tashkent 100149, Uzbekistan}
\affiliation{Urgench State University, Kh. Alimjan Str. 14, Urgench 221100, Uzbekistan}
\affiliation{Shahrisabz State Pedagogical Institute, Shahrisabz Str. 10, Shahrisabz 181301, Uzbekistan}

\begin{abstract}

In this work, we have studied the thermodynamic properties of the Van der Waals black hole in the framework of the nonextensive Kaniadakis entropy. We have shown that the black hole properties, such as the mass and temperature, differ from those obtained by using the the Boltzmann-Gibbs approach. Moreover, the nonextensivity $\kappa$-parameter changes behavior of the Gibbs free energy via introduced thermodynamic instabilities, whereas the emission rate is influenced by $\kappa$ only at low frequencies. Nonetheless, the pressure-volume ($P(V)$) characteristics are found independent of $\kappa$ and the entropy form, unlike in other anti-de Sitter (AdS) black hole models. In summary, presented findings partially support previous arguments of Gohar and Salzano that under certain circumstances all entropic models are equivalent and indistinguishable \cite{gohar2024foundations}.
\end{abstract}

\maketitle

\section{Introduction}

Black holes (BHs) are one of the most intriguing entities in the modern physics as they are crucial for understating fundamental relationship between gravity and quantum world, including unification of gravity, quanta and statistical physics \cite{navarro2005,raju2022lessons}. From this perspective, the fact that BH emits (Hawking) radiation due to the evaporation process, provided pivotal concepts of Hawking temperature and Bekenstein entropy associated with the BH horizon \cite{hawking1974black,bekenstein1973black}. In fact, the process of evaporation appears to be non-unitary, leading to the famous black hole (or information loss) paradox \cite{giddings1995black,raju2022lessons}. In this manner, thermodynamic properties of BH yield serious challenges, even without invoking explicitly quantum field theory. In particularly, maximal information falling into BH can be expressed as entropy \cite{navarro2005,Kubizvnak:2015bh,Carlip:2014pma,Kubiznak:2016qmn,Cong:2021fnf}. Moreover, there are some intriguing aspects regarding entropy of a BH, when one considers the holographic principle as a guiding physical postulate \cite{hooft1993dimensional,susskind1995world}. According to this principle, the black hole horizon can store information in the form of a hologram. In this context, anti-de Sitter (AdS) black holes have been subject of important studies in recent decades, mainly due the so called AdS/CFT correspondence \cite{Witten:1998qj}. The AdS/CFT conjecture states that the gravity sector involving the non-compact AdS space, can be interpreted as a thermal field theory from an asymptotic viewpoint. Another interesting finding  associated with the AdS/CFT is the exploration of first-order phase transitions in black holes \cite{Chamblin:1999tk, Chamblin:1999hg, Niu:2011tb}. These transitions have been found to be analogous to those observed for liquid-gas transitions from thermodynamics and chemistry \cite{Sahay:2010wi, Sahay:2010tx, Kubiznak:2012wp}. In this framework, the negative cosmological constant is interpreted as a pressure \cite{Kastor:2009wy, Dolan:2010ha}, leading to an enhanced phase space where the equation of state $P = P(V, T)$ allows for the study of the critical behavior of AdS black holes \cite{Dolan:2010ha, Dolan:2011xt}. The analogy between black holes and real gases is further extended by the Van der Waals (VdW) black hole model, which incorporates critical behavior and phase transitions similar to those observed in VdW fluids \cite{rajagopal2014}. This model provides a richer thermodynamic description, where the pressure-volume-temperature relationship mimics the VdW equation for real fluids \cite{rajagopal2014}. 

However, standard description of BH entropy linked with the Boltzmann-Gibbs (BG) viewpoint gives rise to rather unconventional scaling. The holographic principle indicates that the BG entropy $S_{BG}$ of a (3+1)-dimensional black hole is proportional to its area $L^2$, rather than its volume $L^3$. Similarly, for strongly quantum-entangled $d$-dimensional systems, the area law states that $S_{BG}$ is proportional to $\ln L$ if $d=1$, and to $L^{d-1}$ if $d>1$, rather than $L^d$ ($d \geq 1$). Consequently, the extensiveness of thermodynamic entropy of a $d$-dimensional system no longer holds true \cite{tsallis2013black}. This issue led to the search for generalizations and nonextensive approaches to the statistical description of thermodynamics, introducing concepts such as Barrow or Tsallis statistics within the BH thermodynamics \cite{Tsallis:2019,ilic2021overview,ladghami2024barrow}. These possibilities of describing statistics in relativity are connected to the thermodynamic conjecture, which treats spacetime as an ordinary thermal system with Einstein equations viewed as a equivalent to the laws of thermodynamics \cite{Jacobson:1995ab,Cai:2006rs,Lymperis:2021qty}. In this context, Kaniadakis statistics (known as $K$-statistics) is of particular interest due being a coherent relativistic statistical theory, as it is inspired by the symmetry of the Lorentz group \cite{Kaniadakis:2001, Kaniadakis:2002zz, Kaniadakis:2005zk, Kaniadakis:2006xn}. Kaniadakis approach have gained significant attention, especially in cosmology, for its applicability to the slow-roll inflation mechanism \cite{Lambiase:2023ryq} and the PeV neutrino tension \cite{Blasone:2023yke}. As for the BH physics, Kaniadakis statistics poses the ability to accurately describe the Bekenstein-Hawking area law \cite{Alonso-Serrano:2020hpb}. Therefore, there are compelling reasons to extend BG statistics for black hole physics, with the Kaniadakis approach being particularly favored for its inherent relativistic essence 
\cite{Kaniadakis:2001,Kaniadakis:2002zz,Kaniadakis:2005zk,Kaniadakis:2006xn}.

%Motivated by the interesting interplay between BHs chemistry concept, non-extensive statistics in BH physics and interaction between BHs and molecules microscopic behavioUr, the following work aims to study main thermodynamic aspects of Van der Waals BH . The outline of the present work is as follows: in Sec~\ref{sec2} we present the significant description of what Van der Waals BHs stands for and exposing some of their thermodyanmic quantities. In Sec~\ref{sec3}, after introducing the needed standard thermodynamic set, we discuss the relevant thermodynamic stability revealing the local stability and exploring the scenario involved the Gibbs free energy. Afetrwards, the examination of the energy emission rate is taken to be as an important task of this paper. Finally, Sec~\ref{con} presents the summary and conclusions of our work.

Motivated by the interesting interplay between BH chemistry concept and non-extensive statistics in BH physics, the following work aims to study main thermodynamic aspects of VdW BH. The outline of the present work is as follows: in Sec. II, we present the description of VdW BH and basic thermodynamic quantities in the $\kappa$-statistics regime. In Sec. III, we discuss the thermodynamic stability revealing the local stability and the Gibbs free energy. After that, the examination of the energy emission rate is carried out in Sec. IV. Finally, Sec. V summarizes and draw conclusions of the work presented herein.

\section{Kaniadakis statistics and Van der Waals black hole geometry}

The starting point in the description of Van der Waals BH is the modified line element of AdS black hole, in the following form: \cite{rajagopal2014,ditta2023thermal}
\begin{align}
    ds^2=-f(r)dt^2+f(r)^{-1}dr^2+r^2d\theta^2+r^2\sin^2(\theta)d\phi^2,
    \label{metric}
\end{align}
where the radial function takes form:
\begin{align}
    f(r)=2\pi a - \frac{2 M}{r}+\frac{r^2}{l^2}(1+\frac{3b}{2r})-\frac{3\pi a b^2}{r(2r+3b)}-4\frac{\pi a b}{r}\log(\frac{r}{b}+\frac{3}{2}),
\end{align}
where magnitude of the intramolecular forces and molecular pressure are $a>0$ and $b>0$, respectively \cite{rajagopal2014,atkins2023atkins}. Notably, for $a=\frac{1}{2\pi}$ and $b=0$ the metric reduces to the usual Schwarzschild AdS line element. One can view the VdW BH as a thermodynamic generalization of the Schwarzschild black hole. Moreover, in that scenario, pressure can be associated with the cosmological constant $\Lambda$ via \cite{rajagopal2014,Kubizvnak:2015bh}:
\begin{align}
    P=\frac{3}{8\pi l^2}=-\frac{\Lambda}{8\pi}.
\end{align}
Thus, metric (\ref{metric}) is the extended version of the AdS BH spacetime. In order to carry out an investigation in the extended phase space, we also introduce thermodynamic volume in the following manner:
\begin{align}
    V=\big(\frac{\partial M}{\partial P} \big)_{S,Q}=\frac{2}{3}r_h^2(3b+2r_h).
\end{align}
Then, mass can be obtained from $f(r)=0$:
\begin{align}
    M=\frac{1}{6}\pi\Big[\frac{a(-9b^2+18 b r_h+12r_h^2)}{3b+2r_h}-12 ab \log(\frac{r_h}{b}+\frac{3}{2})+4P r_h^2(3b+2r_h)\Big].
    \label{massr}
\end{align}
Additionally, it is known that Van der Waals black hole satisfy first law of the BH thermodynamics \cite{rajagopal2014,bronnikov2021}:
\begin{align}
    dM=TdS+\Phi d Q+VdP,
    \label{firstlaw}
\end{align}
as well as the Smarr relation:
\begin{align}
    M=2(TS-VP)+\Phi Q,
\end{align}
where 
\begin{align}
    T=\big(\frac{\partial M}{\partial S}\big)_{P,Q},\;\;\; \Phi=\big(\frac{\partial Q}{\partial S}\big)_{S,P}.
    \label{eq9}
\end{align}
The Bekenstein-Hawking entropy of the VdW BH according to the standard Boltzmann-Gibbs statistics is: 
\begin{align}
    S_{BH}=\frac{A_{BH}}{4}=\pi r_h^2.
    \label{beke}
\end{align}
However, as already mentioned in the introduction, BG statistics may not be an appropriate way to describe the thermodynamic properties of a black hole, proving the need for viable forms of entropy. In what follows, Kaniadakis statistics leads to the relativistic notion of entropy. Please note that the relativistic nature of Kaniadakis statistics is associated with the inclusion of the relativistic relation between energy and momentum, as well as possessing necessary relativistic invariance \cite{Kaniadakis:2001,Kaniadakis:2002zz,Kaniadakis:2005zk,Kaniadakis:2006xn}.

\begin{figure}
    \centering
    \includegraphics{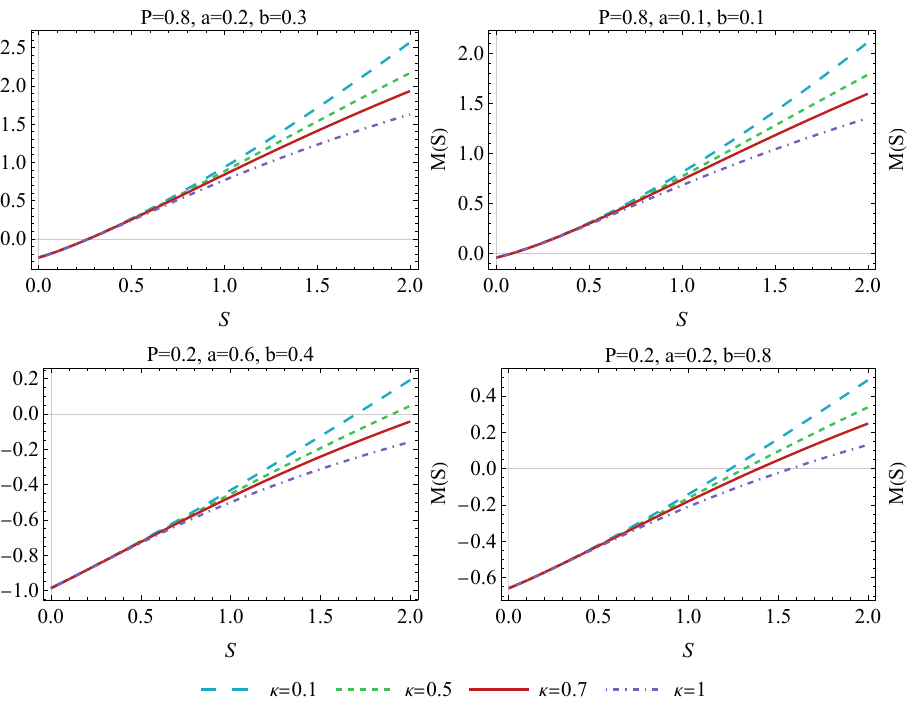}
    \caption{The mass parameter $M$ as a function of entropy $S$ for a different values of $\kappa$ parameter.}
    \label{figm}
\end{figure}

In details, according to the Kaniadakis statistics, the entropy formula (\ref{beke}) is modified and takes the following form \cite{Kaniadakis:2005zk}:
\begin{align}
    S=\frac{1}{\kappa}\sinh(\kappa S_{BH}),
    \label{kan}
\end{align}
being nonextensive in its nature. Note, that modified entropy is a monotonically increasing function of $S_{BH}$ and as a consequence also a function of horizon radii $r_h$. Moreover, for the $\kappa\rightarrow 0$, standard Bekenstein-Hawking entropy is recovered. It is worth to add, that since the function is even, we can focus on the $\kappa>0$ without losing generality \cite{Lymperis:2021qty}. From area law and expression for $\kappa$-entropy, we get an inverse:
\begin{align}
    r_h=\frac{\sqrt{\frac{\sinh ^{-1}(\kappa S)}{\kappa }}}{\sqrt{\pi }},
    \label{rhs}
\end{align}
which will be used to study how Kaniadakis entropic description influences the properties of the Van der Waals black hole solution.
Now, the mass of a black hole (\ref{massr}) can be expressed as a function of entropy ($S$):
\begin{align}\nonumber
    M(S)&=\frac{\sqrt{\pi } a \left(-3 \pi  b^2 \kappa +6 \sqrt{\pi } b \kappa  \sqrt{\frac{\sinh ^{-1}(\kappa S)}{\kappa }}+4 \sinh ^{-1}(\kappa S)\right)}{6 \sqrt{\pi } b \kappa +4 \kappa  \sqrt{\frac{\sinh ^{-1}(\kappa S)}{\kappa }}}-2 \pi  a b \log \left(\frac{\sqrt{\frac{\sinh ^{-1}(\kappa S)}{\kappa }}}{\sqrt{\pi } b}+\frac{3}{2}\right)\\ &+\frac{2 P \sinh ^{-1}(\kappa S) \left(3 b+\frac{2 \sqrt{\frac{\sinh ^{-1}(\kappa S)}{\kappa }}}{\sqrt{\pi }}\right)}{3 \kappa }.
    \label{massm}
\end{align}
This mass function actually contains different limits, leading to the different BH mass relations. For example, in BG limit of Kaniadakis statistics ($\kappa \rightarrow 0$), usual mass-entropy relationship of VdW BH is recovered:
\begin{align}
  M(S)=  \frac{3 \pi  a \left(-3 \pi  b^2+6 \sqrt{\pi } b \sqrt{S}+4 s\right)-12 \pi ^{3/2} a b \left(3 \sqrt{\pi } b+2 \sqrt{S}\right) \log \left(\frac{\sqrt{S}}{\sqrt{\pi } b}+\frac{3}{2}\right)+4 P s \left(3 \sqrt{\pi } b+2 \sqrt{S}\right)^2}{6 \left(3 \pi  b+2 \sqrt{\pi } \sqrt{S}\right)}.
\end{align}
The behaviour of the mass as a function of entropy has been plotted in Fig. (\ref{figm}). Initially, the mass of a BH may be negative and thus unstable. However, once the entropy is increased it will eventually reach the positive values with the rate of increase being affected by the values of $\kappa$, i.e. the lower the $\kappa$ gets this increase is more rapid. While the Kaniadakis entropy has no impact on the early stages it significantly affects the final growth rate of a mass function. Note that similar behavior has been observed for a charged AdS black holes within $\kappa$-statistics approach \cite{Luciano:2023bai}. 

On the other hand, from the first law of BH thermodynamics (\ref{firstlaw}) and Eq. (\ref{massm}), the thermal radiation of a VdW BH reads:
\begin{align}
&T(S)=\\ \nonumber
&\frac{2 \left(\pi ^{3/2} b \kappa ^2 \left(a+9 b^2 P\right) \sqrt{\frac{\sinh ^{-1}(\kappa  S)}{\kappa }}+\sinh ^{-1}(\kappa  S) \left(\pi  a \kappa +21 \pi  b^2 \kappa  P+16 \sqrt{\pi } b \kappa  P \sqrt{\frac{\sinh ^{-1}(\kappa  S)}{\kappa }}\right)+4 P \sinh ^{-1}(\kappa  S)^2\right)}{\kappa ^2 \sqrt{\pi  \kappa ^2 S^2+\pi } \sqrt{\frac{\sinh ^{-1}(\kappa  S)}{\kappa }} \left(3 \sqrt{\pi } b+2 \sqrt{\frac{\sinh ^{-1}(\kappa S)}{\kappa }}\right)^2}.
\label{ts}
\end{align}
%\cite{rajagopal2014}:

%$v=2 r_H + 3b$

\begin{figure}
    \centering
    \includegraphics{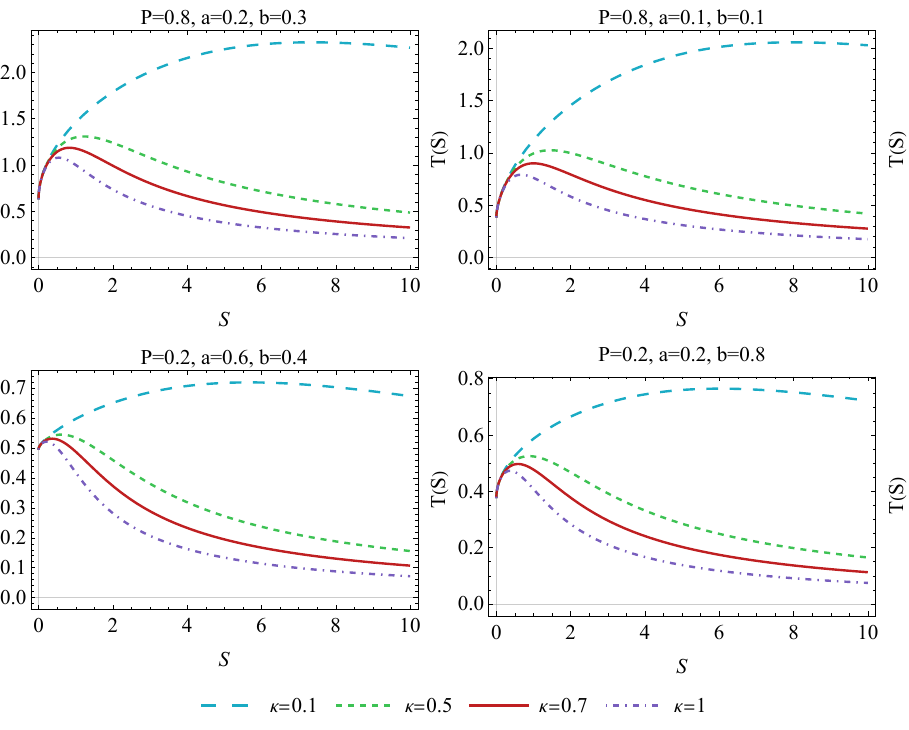}
    \caption{Thermodynamic temperature $T$ as a function of entropy $S$ for a different values of $\kappa$ parameter. The diagrams has been plotted for different values of $P$, $a$ and $b$.}
    \label{fig2}
\end{figure}

The $T(S)$ relation is presented in the Fig. (\ref{fig2}), where different choices of the BH parameters $(a,b,P)$ are taken into account. This graphical analysis reveals that only $\kappa \rightarrow 1$ changes the behavior in a notable way. Interestingly, for some value of $S$, radiation temperature reaches its peak $T_{crit}$. The peak shifts towards higher entropies as Kaniadakis statistics becomes more relevant (i.e. $\kappa$ increases). 

\section{Thermal stability}

\subsection{Heat capacity}

\begin{figure}
    \centering
    \includegraphics{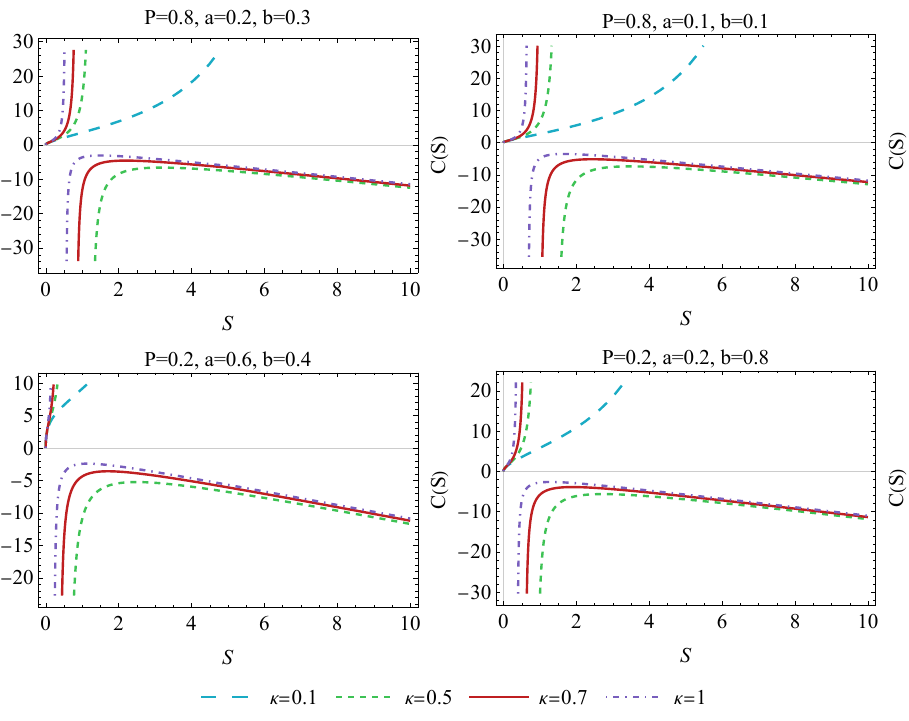}
    \caption{Heat capacity of a Van der Waals black hole as a function of entropy $S$ for a different values of $\kappa$, $P$, $a$ and $b$.}
    \label{fig3}
\end{figure}

In order to investigate the stability of the Van der Waals BH system, and see how Kaniadakis entropy affects it, one needs to analyze its heat capacity $C$. Positive values of $C$ are associated with the stable region, while negative ones can suggest the unstable region. Additionally, potential divergences in the heat capacity may shed light on phase transitions and its interpretation \cite{Kubizvnak:2015bh,Hendi:2015rja,Kubiznak:2016qmn,Ditta:2024iky}.
In what follows, heat capacity of a black hole can be defined as \cite{Hendi:2015rja,Cai:2003kt}:
\begin{align}
    C=\frac{\frac{\partial M}{\partial S}}{\frac{\partial^2 M}{\partial S^2}}=\frac{T}{\frac{\partial^2 M}{\partial S^2}},
\end{align}
and for the Kaniadakis statistics of VdW black hole reads ($C_0=T$):
\begin{align}
  C=\frac{C_0}{C_1+C_2+C_3},
\end{align}
where:
\begin{align}
    C_1&=8 P\kappa ^2  S \sinh ^{-1}(\kappa  S)^2 \left(11 \sqrt{\pi } b+2 \sqrt{\frac{\sinh ^{-1}(\kappa  S)}{\kappa }}\right),\\ \nonumber
    C_2&=\kappa \sinh ^{-1}(\kappa  S) \big(2 \pi  a \kappa ^2 S\left(5 \sqrt{\pi } b+2 \sqrt{\frac{\sinh ^{-1}(\kappa  S)}{\kappa }}\right)+162 \pi ^{3/2} b^3 \kappa ^2 P S+180 \pi  b^2 \kappa ^2 P S \sqrt{\frac{\sinh ^{-1}(\kappa  S)}{\kappa }}\\ &-36 b P \sqrt{\pi  \kappa ^2 S^2+\pi }-8 P \sqrt{\kappa ^2 S^2+1} \sqrt{\frac{\sinh ^{-1}(\kappa S)}{\kappa }}\big),\\ \nonumber
    C_3&=\pi \kappa^2 \Bigg(a \left(6 \pi  b^2 \kappa ^2 s \sqrt{\frac{\sinh ^{-1}(\kappa  s)}{\kappa }}+b \sqrt{\pi  \kappa ^2 s^2+\pi }+2 \sqrt{\kappa ^2 s^2+1} \sqrt{\frac{\sinh ^{-1}(\kappa  s)}{\kappa }}\right)\\  &+27 b^2 P \left(2 \pi  b^2 \kappa ^2 s \sqrt{\frac{\sinh ^{-1}(\kappa  s)}{\kappa }}-b \sqrt{\pi  \kappa ^2 s^2+\pi }-2 \sqrt{\kappa ^2 s^2+1} \sqrt{\frac{\sinh ^{-1}(\kappa  s)}{\kappa }}\right)\Bigg).
\end{align}
The numerator of the expression, $C_0$, consists of terms dependent on $\kappa$, $a$, $b$, and $P$, reflecting the influence of Kaniadakis entropy, intermolecular forces, molecular pressure, and $\Lambda$ on the heat capacity. Additionally, the complexity of the denominator underscores the intricate interplay between these parameters in shaping the heat capacity. The behaviour of the heat capacity under the different values of $\kappa$ is presented in Fig. (\ref{fig3}). Notably, a change of sign corresponds to the presence of divergent and physical limitation points, indicating a phase transition process, either a second-order or a first-order phase transition, respectively. Note that for the certain values of $\kappa$ and $S$, the heat capacity indicates local instability of a VdW BH. On the other hand, for $\kappa \rightarrow 0$ the BH eventually will be stable as $C(S)$ diagram will approach standard characteristics of the VdW BH \cite{rajagopal2014,ditta2023thermal}.

\subsection{$P-V$ criticality}

The analysis of the critical behavior of the VdW BH within the Kaniadakis statistics regime can be done by using the $P-V$ phase transitions. Notably, Van der Waals fluids are characterized by the $P-T$ relationship given by:
\begin{align}
    P+\frac{a}{V^2}(V-b)=T,
\end{align}
or:
\begin{align}
    \frac{-a b+a V-T V^2}{V^2 (b-V)}=P.
    \label{pwdw}
\end{align}
In this part of the work, we will derive analogous equations for the VdW BH in the extended phase space of Kanidakis thermodynamics.

After some algebra, by using (\ref{ts}), (\ref{kan}) and (\ref{beke}), we get the relationship between pressure and the horizon radius:
\begin{align}
    P(r_h)=\frac{T \cosh \left(\pi  \kappa  r_h\right)}{2 (b+r_h)}-\frac{a}{(3 b+2 r_h)^2}.
    \label{prh}
\end{align}
In order to compare Eq.(\ref{prh}) with the VdW relation (\ref{pwdw}) one can associate the horizon radius with the specific volume $V$ \cite{rajagopal2014}:
\begin{align}
    r_h=\frac{V-3}{b}.
\end{align}
Then the $P(V)$ relationship within Kaniadakis framework is:
\begin{align}
    P(V)=-\frac{a}{V^2}-\frac{T \cosh \left(\frac{1}{4} \pi  \kappa (V-3 b)\right)}{b-V},
    \label{pvkan}
\end{align}
and for $\kappa \rightarrow 0$, the behavior of the Van der Waals fluid (\ref{pwdw}) is recovered. 

In order to determine the qualitative behavior of critical points we note that those points occur when $P(v)$ has infection point, i.e. 
\begin{align}
    \frac{\partial P}{\partial V}=0,\;\;\;\; \frac{\partial^2 P}{\partial V^2}=0.
    \label{cond}
\end{align}
Since Eq. (\ref{pwdw}) can be expressed in the cubic form,
\begin{align}
    -a b+a V-b P V^2+P V^3-T V^2=0,
\end{align}
the critical temperature, volume and pressure can be obtained from comparison with the coefficients of the equation $P_c(V-V_c)^3=0$.
To proceed further, we need to expand expression (\ref{pvkan}) according to the $cosh (x)=1+\frac{x^2}{2}+O(...)$. This leads to the following relation:
\begin{align}
    P(V)\approx \frac{-a b+a V-T V^2}{V^2 (b-V)}-\frac{\kappa ^2 \left(\pi ^2 T (V-3 b)^4\right)}{32 (b-V)}+O\left(\kappa ^3\right).
\end{align}
Now, from the $P_c(V-V_c)^3=0$ and conditions (\ref{cond}), we can get the critical coefficients:

\begin{align}
    P_c= \frac{a}{27 b^2},\ \ \ \ V_c=3 b,\ \ \ \ T_c= \frac{8 a}{27 b}.
\end{align}

Interestingly, they do not depend on the constant $\kappa$, coinciding with the coefficients of the Van der Waals fluid \cite{Kubiznak:2012wp}. Hence, when BH behaves explicitly like the VdW fluid, the different entropies lead to the effectively similar behavior in the $P-V$ diagrams as one obtained using Bekenstein and Hawking laws \cite{navarro2005}. Hence, our analysis partially supports arguments on thermodynamic inconsistencies as was mentioned in \cite{gohar2024foundations}. In details, therein it was shown that the choice of entropy does not really matter since results for all entropies coincide with the ones obtained from usual Gibbs-Boltzmann statistics. Since the $P(V)$ characteristic is independent on the form of entropy, for details on the  characteristics for a VdW BH in standard statistical regime we refer to the \cite{rajagopal2014,ditta2023thermal}.

\subsection{Gibb's free energy}

In order to obtain a complete description of the phase transition for the BH system we adopt a special thermodynamic potential. In this way, the Gibbs free energy is a thermodynamic quantity evaluated on the grounds of Euclidean action via a tailored limiting factor. A global stability analysis is carried out on the evidence of the sign of the Gibbs free energy. In the extended phase space, the thermodynamic potential is in turn the Gibbs free energy $G=M-T\, S=H-T\, S$. It is worth noting here that all discontinuous changes in the first- or second-order derivatives of the Gibbs energy induce a first- or second-order phase transition in the process. Hence, the Gibbs free energy can be defined as follows:

\begin{align}\nonumber
 G=&\frac{1}{6} \pi  \left(\frac{a \left(-9 b^2+18 b r_h+12 r_h^2\right)}{3 b+2 r_h}-12 a b \log \left(\frac{r_h}{b}+\frac{3}{2}\right)+4 P r_h^2 (3 b+2 r_h)\right)\\ \nonumber &-\frac{1}{\kappa }\frac{\frac{a \left(-9 b^2+18 b r_h+12 r_h^2\right)}{3 b+2 r_h}-12 a b \log \left(\frac{r_h}{b}+\frac{3}{2}\right)+4 P r_h^2 (3 b+2 r_h)}{12 r_h^2}+\frac{3 a b^2 (3 b+4 r_h)}{4 \left(3 b r_h+2 r_h^2\right)}-\frac{2 a b}{3 b r_h+2 r_h^2}\\ &+a b \log \left(\frac{r_h}{b}+\frac{3}{2}\right)+b P+\frac{4 P r_h}{3}\text{Sinh}\left[\pi  \kappa  r_h^2\right].
\end{align}

\begin{figure}
    \centering
    \includegraphics{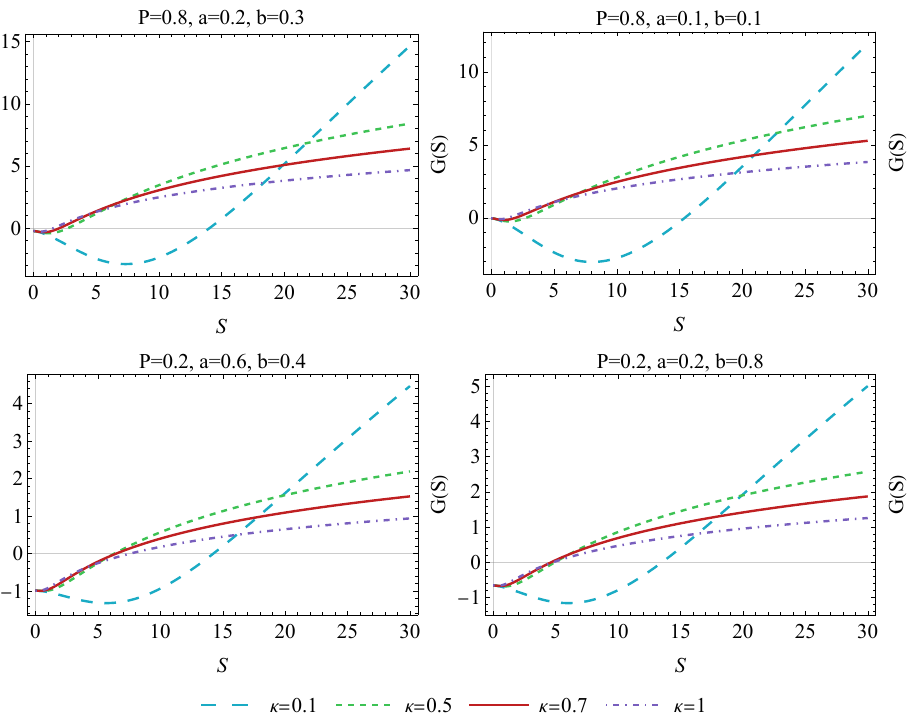}
    \caption{The Gibbs free energy diagram of a Van der Waals black hole as a function of entropy $S$, for different pressures $P$, constants $a$ and $b$ and varying values of of $\kappa$.}
    \label{fig4}
\end{figure}

The functional behavior of the Gibbs free energy in the form of the $G-S$ diagram has been shown in the Fig. (\ref{fig4}). Note, that as $\kappa$ increases the curves become steeper, particularly at higher entropy values. This reflects that larger $\kappa$ values tend to amplify the deviations from the usual thermodynamic behavior \cite{rajagopal2014,Kubiznak:2016qmn,ditta2023thermal}. The analysis reveals that the $G(S)$ in Van der Waals black holes is highly sensitive to the Kaniadakis entropy parameter $\kappa$ and the Van der Waals parameters $a$ and $b$. Negative values of Gibbs energy, particularly at small values of entropy, suggest the potential for phase transitions or thermodynamic instability. This effect is more pronounced for larger values of $a$ and lower pressure. Additionally, as entropy increases, the system generally becomes more stable. However, strong interactions (larger $a$) and low pressure can lead to complex, potentially unstable, thermodynamic states at low entropy.

\section{Energy emission}

Since BH radiates, the quantum fluctuations impose creation of the particles and antiparticles, beyond the event horizon $r_h$. In fact, through quantum tunneling phenomenon, particles with the positive energy may escape the innermost regions of Hawking radiation, giving raise to the evaporation process. The rate at which black hole evaporates is dependent on its energy emission rate. This rate can be observed by an observer located far from the black hole. For such an observer, the black hole's shadow corresponds to a high-energy absorption cross-section. This absorption cross-section characterizes how the black hole interacts with incoming radiation. In what follows, the limiting value of the cross-section is approximately \cite{wei2013observing,ditta2023thermal}:
\begin{align}
    \sigma_{lim} \approx \pi r_h^2.
    \label{eq30}
\end{align}
Thus, the BH energy emission rate is given by:
\begin{align}
    \frac{d^2 \epsilon}{d \omega d t}=\frac{2 \pi^2 \sigma_{lim}}{e^{\omega/T}-1} \omega^3,
    \label{eq31}
\end{align}
for the modified temperature $T$ from Eq. (\ref{ts}). Then by using Eq. (\ref{eq9}) and (\ref{beke}), together with relation (\ref{eq30}), one can get emission rate as a function of $r_h$ and VdW parameters. Hence, final  expression for the energy emission is:
\begin{align}\nonumber 
  \frac{d^2 \epsilon}{d \omega d t}&= \frac{2\left(\pi ^{3/2} b \kappa ^2 \left(a+9 b^2 P\right) \sqrt{\frac{\sinh ^{-1}\left(\sinh \left(\pi  \kappa  \text{rh}^2\right)\right)}{\kappa }}+4 P \sinh ^{-1}\left(\sinh \left(\pi  \kappa  \text{rh}^2\right)\right)^2+\sinh ^{-1}\left(\sinh \left(\pi  \kappa  \text{rh}^2\right)\right)\right)}{\kappa ^2 \sqrt{\frac{\sinh ^{-1}\left(\sinh \left(\pi  \kappa  \text{rh}^2\right)\right)}{\kappa }} \sqrt{\pi  \sinh ^2\left(\pi  \kappa  \text{rh}^2\right)+\pi } \left(3 \sqrt{\pi } b+2 \sqrt{\frac{\sinh ^{-1}\left(\sinh \left(\pi  \kappa  \text{rh}^2\right)\right)}{\kappa }}\right)^2}\\ &\times \pi  a \kappa +21 \pi  b^2 \kappa  P+16 \sqrt{\pi } b \kappa  P \sqrt{\frac{\sinh ^{-1}\left(\sinh \left(\pi  \kappa  \text{rh}^2\right)\right)}{\kappa }}.
\end{align}

\begin{figure}[!htb]
   \begin{minipage}{0.4\textwidth}
     \centering
     \includegraphics[width=1.
     \linewidth]{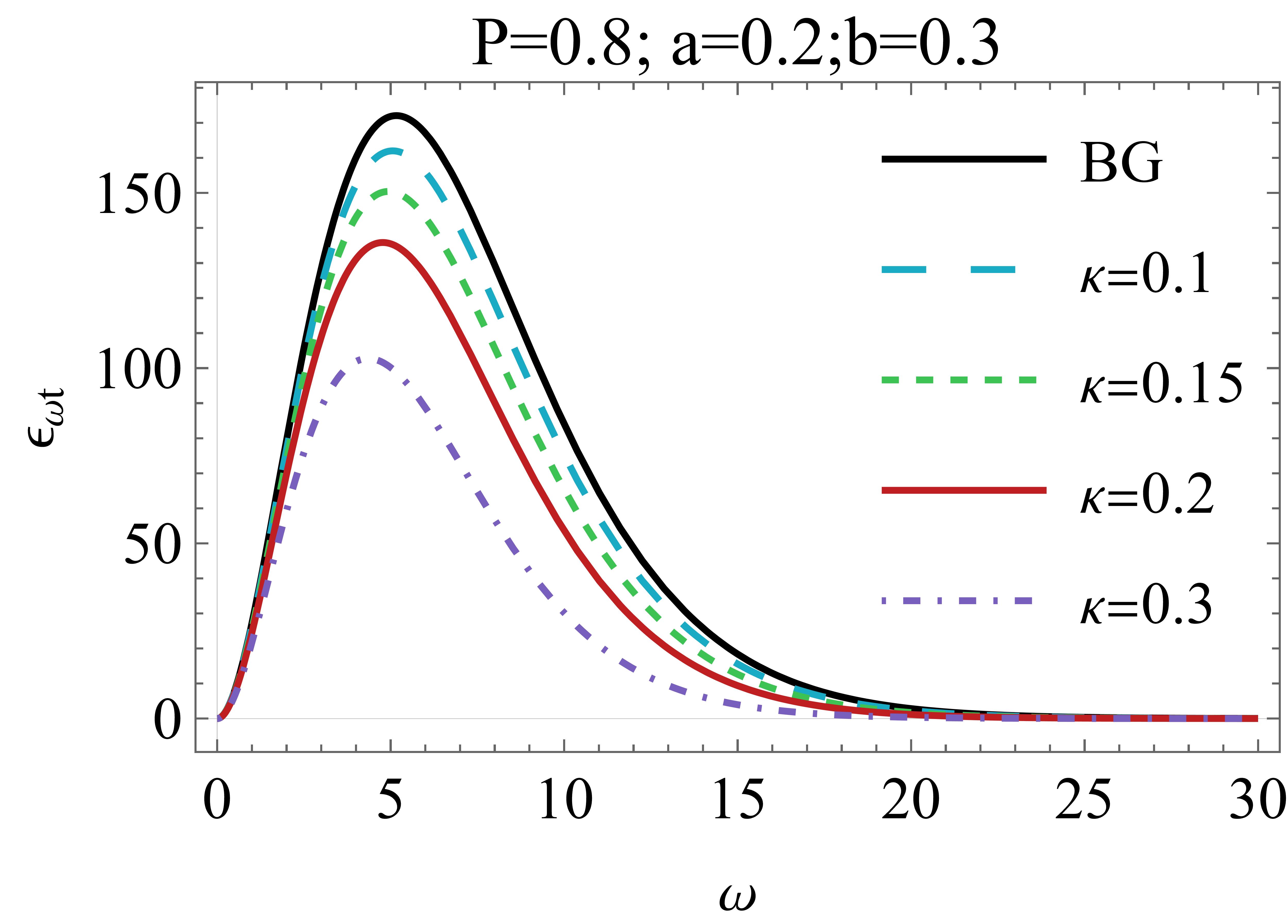}
   \end{minipage}\hfill
   \begin{minipage}{0.4\textwidth}
     \centering
     \includegraphics[width=1.\linewidth]{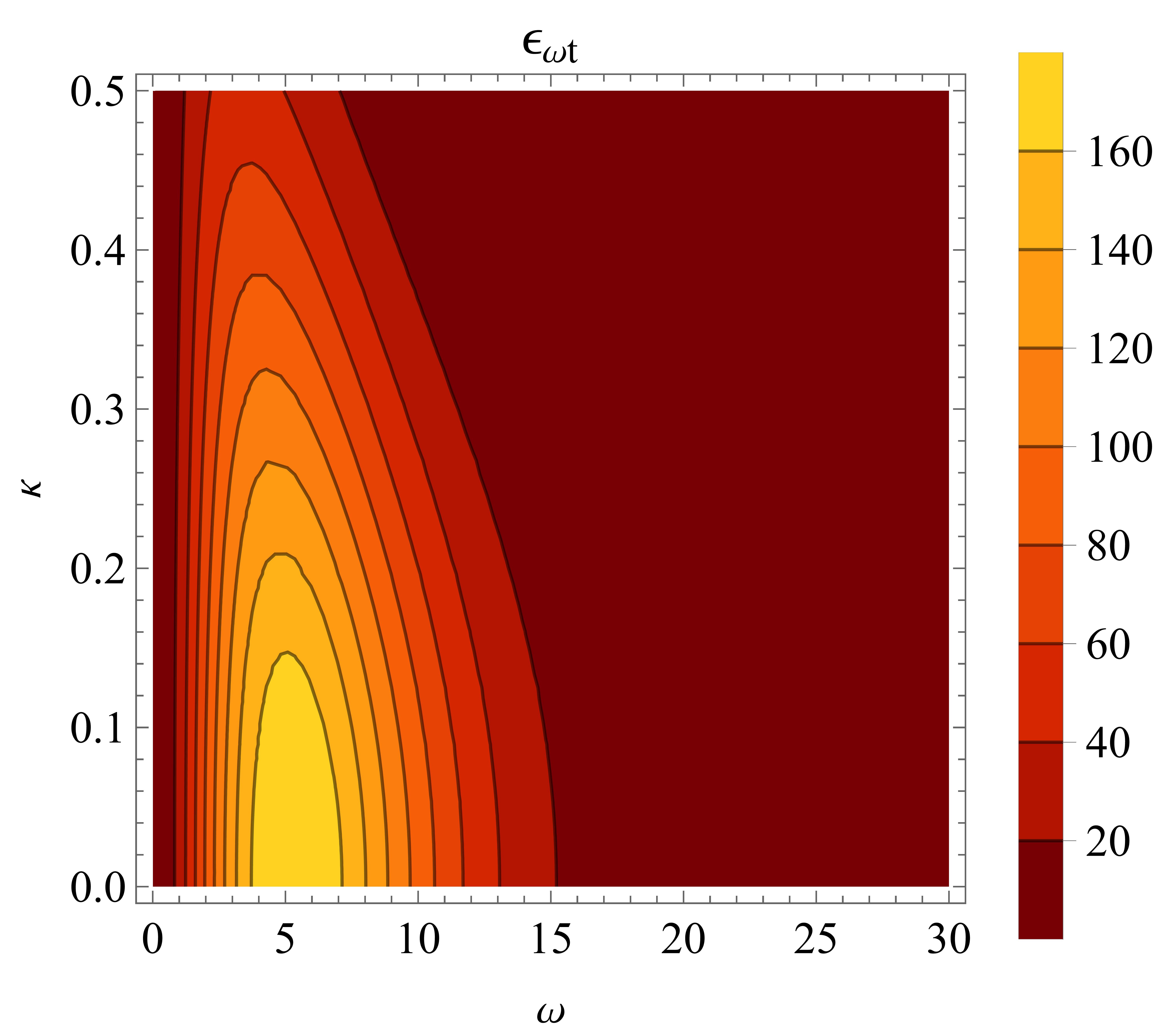}
\end{minipage}
    \caption{The energy emission rate as a function of $\omega$ for different choices of free parameters and $\kappa$. The black line corresponds to the standard entropy from Boltzmann-Gibbs (BG) statistics.}
     \label{fig:5}
\end{figure}

The behavior of the energy emission rate as a function of $\omega$ for different choices of $\kappa$ is presented in the Fig. (\ref{fig:5}). One can see that there exists a peak of the energy emission rate for the VdW black hole. Clearly, the introduction of the Kaniadakis entropy dampens the emission rate and lowers this peak. This means that under nonextensive statistics, at least at the begining of the process, black hole evaporates in a slower pace. However, as the $\omega$ increases, the effect of the $\kappa$-statistics is indistinguishable from the behavior associated with the Boltzmann-Gibbs formula. Hence, at the regimes of the high $\omega$, the choice of the entropy (\ref{kan}) is not significant.  

%\begin{figure}
 %   \centering
  %  \includegraphics{figem.pdf}
   % \caption{The energy emission rate as a function of $\omega$ for different choices of free parameters and $\kappa$.}
   % \label{fig5}
%\end{figure}
%\section{Conclusion}\label{sec8}
%\label{conc}

%In this research paper, we have explored the concept of Kaniadakis statistics in conjunction with the investigation of thermodynamics of the Van der Waals BH \cite{rajagopal2014}.

%It is important to add some comments on $P(v)$ the relationship, which is exactly the same as the Van der Waals fluid known from the ordinary thermodynamics.
% Quoting the  "entropic cosmology inconsistency" argument of \cite{gohar2024foundations}: \textit{with consistent thermodynamic quantities defined on the Hubble horizon, satisfying Clausius relation, and employing a linear $M-L$ relation, all the investigated nonextensive entropic force models are identical to the original entropic force models that are derived from the standard Bekenstein entropy and the Hawking temperature}, it is possible that Van der Waals BH may be treated as another testing point of a consistency of nonextensive entropies. In what follows, the role of a Clausius relation for a VdW BH within nonextensive regimes is yet to be analysed.

\section{Summary and Conclusions}

In this work, we investigated the thermodynamics of the Van der Waals black hole within the framework of nonextensive Kaniadakis entropy. Our analysis reveals that the Kaniadakis parameter \(\kappa\) significantly alters the thermodynamic properties of the black hole, particularly its mass, temperature and Gibbs free energy as functions of entropy. These modifications suggest the presence of thermodynamic instabilities at low entropy, with parameters \(a\) and \(b\) of the Van der Waals model further influencing these effects.

Interestingly, our findings show that the pressure-volume $P-V$ behaviour of the Van der Waals black hole remains independent of the chosen entropy form, aligning with the results obtained before by Ditta and colleagues in \cite{ditta2023thermal}. This observation aligns with previous research suggesting that, under specific conditions, nonextensive entropic models are equivalent to those based on standard Boltzmann-Gibbs thermodynamics \cite{gohar2024foundations}. Thus, the form of entropy appears to have minimal impact on the critical behavior of the system. Additionally, we found that Kaniadakis entropy affects the energy emission rate of the black hole, with $\kappa$-parameter leading to a slower evaporation rate during the early stages of the process. However, at higher energies, the emission rate converges with predictions obtained from Boltzmann-Gibbs statistics \cite{ditta2023thermal}, indicating that the influence of Kaniadakis entropy is only significant in low-energy regimes.

Overall, our results constrain the flexibility of the Van der Waals black hole model in accommodating nonextensive statistical frameworks, since some BH properties, such as $P-V$ relationship or energy emission rates are preserved. This study paves the way for further investigation on the role of nonextensive entropies in black hole thermodynamics and their implications for understanding black holes. Referring to the "entropic cosmology inconsistency" argument from \cite{gohar2024foundations}, which states: \textit{with consistent thermodynamic quantities defined on the Hubble horizon, satisfying the Clausius relation, and employing a linear $M-L$ relation, all nonextensive entropic force models are identical to those derived from the standard Bekenstein entropy and Hawking temperature}, the Van der Waals black hole can serve as a valuable testing ground for the nonextensive entropies, such as Kaniadakis, Renyi, Tsallis etc. \cite{ilic2021overview}. Thus, future work should focus on exploration of the phase transitions and the Clausius relation in different statistical regimes for VdW BH, potentially gaining more insights on the consistency (or the lack thereof) nonextensive entropies \cite{ccimdiker2023equilibrium}.

\bibliographystyle{apsrev}
\bibliography{nejlt}
\end{document}